\documentclass[%
 aip,
 amsmath,amssymb,
 reprint,%
]{revtex4-1}

\usepackage{graphicx}% Include figure files
\usepackage{dcolumn}% Align table columns on decimal point
\usepackage{bm}% bold math
\usepackage[utf8]{inputenc}
\usepackage[T1]{fontenc}
\usepackage{mathptmx}
\usepackage{etoolbox}

\usepackage{color}

\makeatletter
\def\@email#1#2{%
 \endgroup
 \patchcmd{\titleblock@produce}
  {\frontmatter@RRAPformat}
  {\frontmatter@RRAPformat{\produce@RRAP{*#1\href{mailto:#2}{#2}}}\frontmatter@RRAPformat}
  {}{}
}%
\makeatother
\begin{document}

\preprint{AIP/123-QED}

\title{An emergence of chiral helimagnetism or ferromagnetism governed by Cr intercalation in a dichalcogenide CrNb$_{3}$S$_{6}$}
% Force line breaks with \\

\author{Y. Kousaka}
\affiliation{Department of Physics and Electronics, Osaka Metropolitan University, Sakai, Osaka 599-8531, Japan}
\email{koyu@omu.ac.jp}

\author{T. Ogura}
\affiliation{Department of Physics and Mathematics, Aoyama-Gakuin University, Sagamihara, Kanagawa 252-5258, Japan}

\author{J. Jiang}
\affiliation{Department of Physics and Electronics, Osaka Metropolitan University, Sakai, Osaka 599-8531, Japan}

\author{K. Mizutani}
\affiliation{Department of Physics and Electronics, Osaka Metropolitan University, Sakai, Osaka 599-8531, Japan}

\author{S. Iwasaki}
\affiliation{Research Institute for Interdisciplinary Science, Okayama University, Okayama, Okayama 700-8530, Japan}

\author{J. Akimitsu}
\affiliation{Research Institute for Interdisciplinary Science, Okayama University, Okayama, Okayama 700-8530, Japan}

\author{Y. Togawa}
\affiliation{Department of Physics and Electronics, Osaka Metropolitan University, Sakai, Osaka 599-8531, Japan}

%\date{\today}% It is always \today, today,
             %  but any date may be explicitly specified

\begin{abstract}
A synthesis of single crystals of chiral dichalcogenides $TM_{3}X_{6}$ ($T$: 3$d$ transition metal, $M$: Nb or Ta, $X$: S or Se) remains an intriguing issue for the investigation of emergent quantum properties such as chiral helimagnetism.
In this study, we investigated a correlation between the quantity of Cr intercalation $x$ and magnetic property in single crystals of a chromium (Cr) intercalated chiral disulfide Cr$_x$Nb$_3$S$_6$ in order to optimize the synthesis condition for the intercalation-controlled single crystals.
The magnetic properties including a magnetic transition temperature $T_{\rm c}$ take different values depending on the samples.
We  systematically grew single crystals of Cr$_{x}$Nb$_{3}$S$_{6}$ with $x$ ranged from 0.89 to 1.03 and found that the amount of the Cr intercalation $x$ is an essential factor in controlling the magnetic properties of the grown crystals.
The magnetization anomaly, which appears in the temperature dependence as evidence of the formation of chiral magnetic soliton lattice (CSL), was observed only in a narrow region of $x$ from 0.98 to 1.03.
The single crystals with $x$ being 0.98 and 0.99 showed the CSL behavior with the highest $T_{\rm c}$ of 133 K.
These results indicate that small amount of defects on the sites for $T$ ions dramatically affects the quality of the single crystals in the synthesis of $TM_{3}$S$_{6}$.
We also discuss an importance of synthesizing enantiopure single crystals of chiral dichalgogenides in order to observe chiral physical properties unique to chiral compounds such as magneto-chiral effect and chiral-induced spin selectivity.
\end{abstract}

\maketitle

\section{Introduction}
A chiral helimagnetic material exhibits a helical magnetic structure with a single handedness. Its formation stems from a competition between symmetric Heisenberg and antisymmetric Dzyaloshinskii-Moriya (DM) exchange interactions. The antisymmetric nature of the DM interaction is strongly coupled to a chiral crystalline structure of chiral helimagnets\cite{Dzyaloshinskii1958, Moriya1960}.
The chiral helimagnets have attracted much attention because of an emergence of nontrivial chiral magnetic textures such as chiral magnetic soliton lattice (CSL)\cite{Dzyaloshinskii1964,Dzyaloshinskii1965a,Dzyaloshinskii1965b,Izyumov1984,Kishine2005}
and chiral magnetic vortices called magnetic Skyrmions \cite{Bogdanov1989,Bogdanov1994}.
The presence of such chiral magnetic order unique to the chiral helimagnets has been observed via neutron scattering or electron microscopy in a recent decade \cite{Muhlbauer2009,Yu2010,Togawa2012}.

\begin{figure}[tb]
\begin{center}
\includegraphics[width=8.5cm]{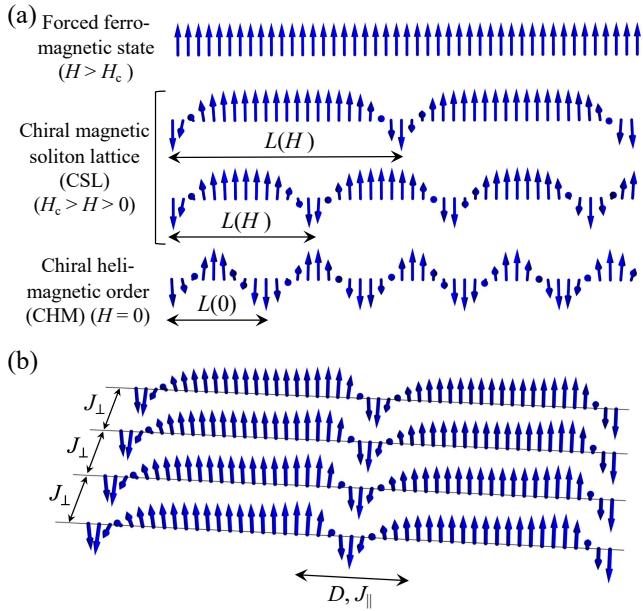}
\end{center}
\caption{Schematic pictures of chiral magnetic order formed in a chiral helimagnet. 
(a) Chiral helimagnetic (CHM) order at zero magnetic field and chiral magnetic soliton lattice (CSL) under a magnetic field applied in a direction perpendicular to the helical axis, followed by a forced ferromagnetic state above a critical field $H_{\rm c}$.
(b) The chiral soliton chains coupled via an in-plane exchange interaction $J_{\perp}$, leading to the formation of robust CSL.
}
\label{f-CSL}
\end{figure}

As envisioned by Dzyaloshinskii, the CSL is a superlattice of chiral spin twists with a uniform periodicity, as shown in Fig.~\ref{f-CSL}(a). Under an external magnetic field applied in the direction perpendicular to the helical axis (principal $c$-axis of the crystal), the harmonic chiral helimagnetic (CHM) order at zero magnetic field transforms into the nonlinear CSL.
Importantly, the CSL period can be controlled continuously by a strength of the magnetic field. Another feature is that the CSL exhibits a robust phase coherence at macroscopic length scale \cite{Togawa2012}.
Because of these intriguing characteristics, the CSL shows a variety of interesting physical responses as exemplified by 
giant magnetoresistance (MR) due to a proliferation of magnetic solitons
\cite{Togawa2013},
discretization effect of MR \cite{Togawa2015}, the presence of surface barrier for the soliton penetration \cite{Yonemura2017}, collective resonant dynamics up to a frequency of sub-terahertz \cite{Shimamoto2022}.
Such coherent, topological, and collective nature of the CSL could be utilized for spintronic device applications \cite{Kishine2015, Togawa2016}.

The number of chiral helimagnets hosting the CSL are still limited. One reason comes from a difficulty in synthesizing helimagnetic materials with a suitable chiral crystalline structure. Another reason could be found in how to detect the CSL.
The CHM period at zero magnetic field $L(0)$ is determined by the ratio of Heisenberg exchange interaction $J_{\parallel}$ and DM interaction $D$ along the helical axis. $L(0)$ is typically about tens of nanometers or longer, which corresponds to some tenth nanometers inverse or less in a wave-vector $k$ space.
Unfortunately, the $Q$ resolution of neutron scattering experiments using a thermal neutron source is not high enough to separate fundamental Bragg and magnetic satellite peaks. Apparently, the CSL has a longer period than that for the CHM and thus the CSL detection using neutron scattering technique becomes much harder. In this connection, there has been a possibility that some compounds of CHM order might be misinterpreted as that of ferromagnetic order in previous studies.

Transition-metal dichalcogenides are one of the appropriate materials that could host the CSL.
More precisely, an intercalation of 3$d$ transition metal ion $T$ into a mother compound $MX_{2}$ ($M:$ Nb or Ta, $X:$ S or Se), provides a rich variety of magnetic properties.
In the case of disulfide, it is known that, depending on a quantity of the intercalated $T$ ions, two different compounds $TM_{4}$S$_{8}$ and $TM_{3}$S$_{6}$ are synthesized.
For clarifying the connection of mother materials, these compounds are sometimes described as $T_{1/4}M$S$_{2}$ and $T_{1/3}M$S$_{2}$, respectively. 
The former has an achiral crystalline structure, while the latter has a chiral crystalline structure with a space group $P6_{3}22$. In the literature before 1980s, it was reported that $TM_{3}$S$_{6}$ showed a variety of magnetic behaviors; paramagnetism for TiNb$_{3}$S$_{6}$, VNb$_{3}$S$_{6}$, TiTa$_{3}$S$_{6}$ and VTa$_{3}$S$_{6}$, antiferromagnetism for FeNb$_{3}$S$_{6}$, CoNb$_{3}$S$_{6}$, NiNb$_{3}$S$_{6}$, CoTa$_{3}$S$_{6}$ and NiTa$_{3}$S$_{6}$, and ferromagnetism for CrNb$_{3}$S$_{6}$, MnNb$_{3}$S$_{6}$, CrTa$_{3}$S$_{6}$, MnTa$_{3}$S$_{6}$ and FeTa$_{3}$S$_{6}$ \cite{Berg1968,Anzenhofer1970,Hulliger1970,Laar1971,Parkin1980a,Parkin1980b}.
In the present study, we mainly focus on CrNb$_{3}$S$_{6}$, which is now known as one of the representative chiral helimagnets for hosting the CSL.
Interestingly, other chiral physical properties unique to chiral materials such as electrical magneto-chiral effect \cite{Aoki2019} and chiral-induced spin selectivity (CISS) \cite{Inui2020} are discussed in CrNb$_{3}$S$_{6}$.

\begin{figure}[tb]
\begin{center}
\includegraphics[width=8.5cm]{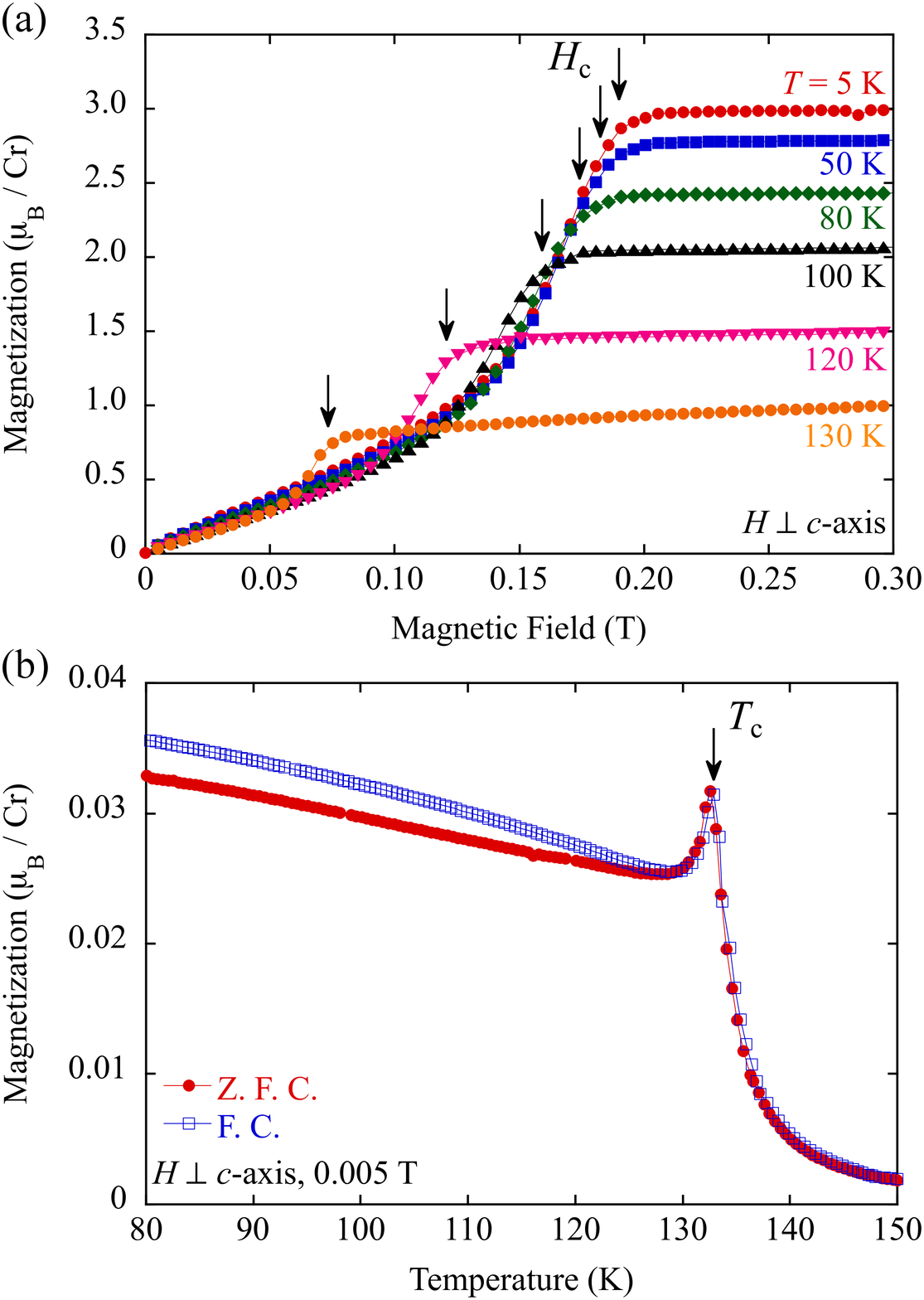}
\end{center}
\caption{
Magnetization curves as functions of magnetic field (a) and temperature (b). The data was taken in the presence of magnetic fields applied perpendicular to the helical axis with the CrNb$_{3}$S$_{6}$ single crystal that shows the maximum $T_{\rm c}$ (133 K) in the present study. Arrows in (a) and (b) represent $H_{\rm c}$ and $T_{\rm c}$, respectively.}
\label{f-CSL-Magnetization}
\end{figure}

In earlier studies around 1970s, CrNb$_{3}$S$_{6}$ was reported to show the ferromagnetic order \cite{Hulliger1970,Laar1971,Parkin1980a,Parkin1980b}.
In 1980s, Miyadai and Moriya investigated a magnetization process in terms of the formation of magnetic helix\cite{Moriya1982}.
They pointed out that it could be interpreted as a formation of the CHM order in CrNb$_{3}$S$_{6}$ and directly observed a diffractive satellite peak that corresponds to the CHM structure with 48 nm by means of small angle neutron scattering \cite{Miyadai1983}.
However, at that time, the field dependence of magnetization was regarded as a discontinuous phase transition between the CHM phase and forced ferromagnetic one without taking any intermediate state. Precisely speaking, the magnetization curve changes continuously toward a critical magnetic field $H_{\rm c}$, as shown in Fig.~\ref{f-CSL-Magnetization}(a).
Such a continuous change of the magnetization should be interpreted as a scenario of the CSL formation, which has been properly discussed in 2000s \cite{Kishine2005,Kousaka2007,Kousaka2009}.
Eventually, the CSL was directly observed by using Lorentz microscopy \cite{Togawa2012}.

Direct observation of the CSL revealed unexpected characteristics of the CSL: the CSL is robust and stable with macroscopic phase coherence.
To realize this property, one-dimensional chains of the chiral twists are strongly correlated with each other, as shown in Fig.~\ref{f-CSL}(b).
The pitch of the chiral twists is governed by the ratio of $J_{\parallel}$ and $D$ along the $c$-axis,
while the coherency arises from an in-plane ($ab$-plane) ferromagnetic exchange interaction $J_{\perp}$.
In CrNb$_{3}$S$_{6}$, the strength of $J_{\perp}$ is as large as the magnetic transition temperature $T_{\rm c}$ and much larger than $J_{\parallel}$\cite{Shinozaki2016}.
Therefore, the value of $T_{\rm c}$ could be a good indicator of the phase coherency of the CSL.

Although $TM_{3}$S$_{6}$ has been investigated for decades, it is still difficult to control the quantity of intercalated $T$.
In the case of CrNb$_{3}$S$_{6}$, the physical properties including $T_{\rm c}$ take different values depending on the samples, as shown in Table~\ref{tbl-SampleComparison}.
For an example, Parkin reported the ferromagnetic order within the basal plane based on the magnetization of CrNb$_{3}$S$_{6}$ with $T_{\rm c}$ = 120 K \cite{Parkin1980a}.
Miyadai found the chiral helimagnetism of CrNb$_{3}$S$_{6}$ with $T_{\rm c}$ = 127 K \cite{Miyadai1983}.

When the CrNb$_{3}$S$_{6}$ sample shows the chiral helimagnetism, the magnetization curve shows a sharp peak as a function of temperature, as shown in Fig.~\ref{f-CSL-Magnetization}(b).
However, a monotonous change of the magnetization appears alternatively in the samples of ferromagnetism as reported in the literature \cite{Parkin1980a,Dyadkin2015,Han2017,Hall2022,Mao2022}.
It would be important to examine the sample quality in terms of the quantity of the Cr ions. Some studies evaluated the quantity of defects in the Cr sites by means of quantitative analysis using single crystal X-ray diffraction \cite{Dyadkin2015, Hall2022} and electron probe microanalysis (EPMA) \cite{Han2017,Mao2022}.
However, there has been no systematic investigation about a correlation between the defects and the magnetic behavior.

\begin{table}[htbp]
      \caption{A list of physical properties of Cr$_{x}$Nb$_{3}$S$_{6}$ single crystals in the literature.
      A magnetic transition temperature $T_{\rm c}$ and a type of magnetism are summarized together with an amount of Cr intercalation $x$, which was determined by electron probe microanalysis (EPMA) or single crystal X-ray diffraction.
      The ferromagnetic or CSL behaviors are categorized by the appearance of the magnetization anomaly in the temperature dependence of magnetization, as shown in Fig.~\ref{f-CSL-Magnetization}(b).
      }
      \label{tbl-SampleComparison}

  \begin{center}

      \begin{tabular}{rrrrr}
      \hline
      \multicolumn{1}{c}{$T_{\rm c}$} [K] & \multicolumn{1}{c}{Magnetism} & \multicolumn{1}{c}{Analysis method} & \multicolumn{1}{c}{$x$} & \multicolumn{1}{c}{Reference}\\
      \hline
      \multicolumn{1}{c}{120} & \multicolumn{1}{c}{Ferro} & \multicolumn{1}{c}{$-$} & \multicolumn{1}{c}{$-$} & \multicolumn{1}{c}{Parkin \it{et al.}~\cite{Parkin1980a}}\\
      \multicolumn{1}{c}{127} & \multicolumn{1}{c}{CSL} & \multicolumn{1}{c}{$-$} & \multicolumn{1}{c}{$-$} & \multicolumn{1}{c}{Miyadai \it{et al.}~\cite{Miyadai1983}}\\
      \multicolumn{1}{c}{128} & \multicolumn{1}{c}{CSL} & \multicolumn{1}{c}{$-$} & \multicolumn{1}{c}{$-$} & \multicolumn{1}{c}{Kousaka \it{et al.}~\cite{Kousaka2009}}\\
      \multicolumn{1}{c}{127} & \multicolumn{1}{c}{CSL} & \multicolumn{1}{c}{$-$} & \multicolumn{1}{c}{$-$} & \multicolumn{1}{c}{Togawa \it{et al.}~\cite{Togawa2012}}\\
      \multicolumn{1}{c}{122} & \multicolumn{1}{c}{CSL} & \multicolumn{1}{c}{$-$} & \multicolumn{1}{c}{$-$} & \multicolumn{1}{c}{Ghimire \it{et al.}~\cite{Ghimire2013}}\\
      \multicolumn{1}{c}{132} & \multicolumn{1}{c}{CSL} & \multicolumn{1}{c}{$-$} & \multicolumn{1}{c}{$-$} & \multicolumn{1}{c}{Togawa \it{et al.}~\cite{Togawa2013}}\\
      \multicolumn{1}{c}{92} & \multicolumn{1}{c}{Ferro} & \multicolumn{1}{c}{X-ray diffraction} & \multicolumn{1}{c}{0.942} & \multicolumn{1}{c}{Dyadkin \it{et al.}~\cite{Dyadkin2015}}\\
      \multicolumn{1}{c}{132} & \multicolumn{1}{c}{CSL} & \multicolumn{1}{c}{$-$} & \multicolumn{1}{c}{$-$} & \multicolumn{1}{c}{Clements \it{et al.}~\cite{Clements2017}}\\
      \multicolumn{1}{c}{125} & \multicolumn{1}{c}{Ferro} & \multicolumn{1}{c}{EPMA} & \multicolumn{1}{c}{0.875} & \multicolumn{1}{c}{Han \it{et al.}~\cite{Han2017}}\\
      \multicolumn{1}{c}{123} & \multicolumn{1}{c}{CSL} & \multicolumn{1}{c}{$-$} & \multicolumn{1}{c}{$-$} & \multicolumn{1}{c}{Togawa \it{et al.}~\cite{Aoki2019}}\\
      \multicolumn{1}{c}{118} & \multicolumn{1}{c}{Ferro} & \multicolumn{1}{c}{X-ray diffraction} & \multicolumn{1}{c}{1.000 (3)} & \multicolumn{1}{c}{Hall \it{et al.}~\cite{Hall2022}}\\
      \multicolumn{1}{c}{56} & \multicolumn{1}{c}{Ferro} & \multicolumn{1}{c}{EPMA} & \multicolumn{1}{c}{0.99 (12)} & \multicolumn{1}{c}{Mao \it{et al.}~\cite{Mao2022}}\\
      \multicolumn{1}{c}{133} & \multicolumn{1}{c}{CSL} & \multicolumn{1}{c}{EPMA} & \multicolumn{1}{c}{0.982 (1)} & \multicolumn{1}{c}{This work}\\
      \multicolumn{1}{c}{133} & \multicolumn{1}{c}{CSL} & \multicolumn{1}{c}{EPMA} & \multicolumn{1}{c}{0.992 (2)} & \multicolumn{1}{c}{This work}\\
      \hline
      \end{tabular}

  \end {center}
\end{table}

The importance of controlling the quantity of intercalated $T$ can be seen in another example of FeTa$_{3}$S$_{6}$ and MnNb$_{3}$S$_{6}$. It is known that they show the ferromagnetic order at around 40 K \cite{Parkin1980a}. However, in some papers\cite{Rahman2022, Hall2022},  
a coexistence of additional magnetic order was reported to occur at 100 K and 160 K in Fe and Mn disulfide compounds, respectively.
These values are likely to correspond to the transition temperatures for MnNb$_{4}$S$_{8}$ \cite{Onuki1986} and FeTa$_{4}$S$_{8}$\cite{Eibschutz1998}.
The studies by using the crystals with a smaller $T_{\rm c}$ and/or those with an impurity phase of $TM_{4}$S$_{8}$ may bring confusing results.

In this paper, we examine how the quantity of the Cr intercalation influences the magnetic property of single crystals of CrNb$_{3}$S$_{6}$ in order to establish a method of growing the intercalation-controlled single crystals. Indeed, we found that the amount of Cr intercalation was an essential factor in controlling the quality of crystals.
In the experiments, we grew single crystals of Cr$_{x}$Nb$_{3}$S$_{6}$ with $x = 0.89 \sim 1.03$ and investigated a correlation between the amount of Cr intercalation and the magnetic property.
The magnetization anomaly, which appears in the temperature dependence as evidence of the CSL formation, was observed only in a narrow region of $x = 0.98 \sim 1.03$.
These results mean that even a small amount of defects in $T$ sites in $TM_{3}$S$_{6}$ will affect the quality of the single crystals significantly.

\section{Crystal Structure}

\begin{figure}[tb]
\begin{center}
\includegraphics[width=8.5cm]{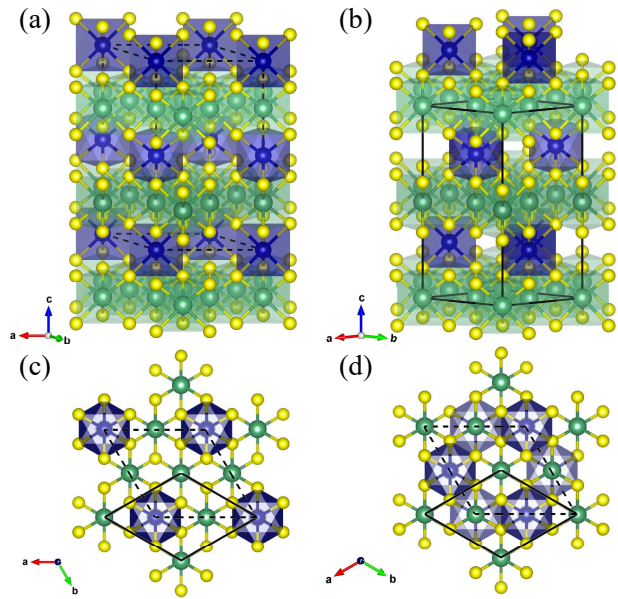}
\end{center}
\caption{
Crystal structures of $3d$ transition metal intercalated niobium disulfides.
The structures of (a) CrNb$_{4}$S$_{8}$ and (b) CrNb$_{3}$S$_{6}$ are respectively illustrated in the side view in (a) and (b), while the ones along the $c$-axis are given in (c) and (d).
Blue, green and yellow balls represent Cr, Nb and S atoms, respectively.
The unit cell of CrNb$_{3}$S$_{6}$ is given by the solid line, while the one of CrNb$_{4}$S$_{8}$ is drawn by the dashed line.
}
\label{f-str}
\end{figure}

The crystal structures of intercalated system of $2H$-$M$S$_{2}$ depend on the intercalation amount of a $3d$ transition metal \cite{Berg1968,Anzenhofer1970,Laar1971}, as shown in Fig.~\ref{f-str}.
CrNb$_{4}$S$_{8}$ forms a centrosymmetric crystal structure with a space group of $P6_{3}/mmc$.  
It has one Cr atom per four units of NbS$_{2}$ and the unit cell along the $a$-axis is twice than that of $2H$-NbS$_{2}$, as seen in Fig.~\ref{f-str}(c).
The location of Cr atoms differs in CrNb$_{3}$S$_{6}$, which forms a chiral crystal structure with a space group of $P6_{3}22$. It has one Cr atom per three units of NbS$_{2}$, while the unit cell along the $a$-axis is $\sqrt{3}$ times longer than that of $2H$-NbS$_{2}$, as shown in Fig.~\ref{f-str}(d).

In CrNb$_{3}$S$_{6}$, the Cr ion is normally intercalated at $2c$ (1/3, 2/3, 1/4) Wyckoff position.
However, there are reports\cite{Berg1968,Laar1971} that the Cr ions take the positions of not only $2c$, but also $2b$ (0, 0, 0) and $2d$ (2/3, 1/3, 1/4).
In the case of CrNb$_{3}$Se$_{6}$, which has the same structure as that of CrNb$_{3}$S$_{6}$, only 5 percent of Cr occupy Wyckoff $2c$ position, but 90 percent occupy Wyckoff $2b$ (0, 0, 0) position \cite{Gubkin2016}.
Resultantly, the magnetic structure is not CHM, but ferromagnetic one.
Similar discussion may hold in CrNb$_{3}$S$_{6}$ and MnNb$_{3}$S$_{6}$ with smaller $T_{\rm c}$.
In particular, when CrNb$_{3}$S$_{6}$ and MnNb$_{3}$S$_{6}$ samples exhibit ferromagnetic behavior in the magnetization without any imprint of the CSL formation,
$3d$ transition metal ions do not fully occupy $2c$ position \cite{Dyadkin2015,Hall2022}.

\section{Experimental Methods}

The polycrystalline samples of Cr$_{x}$Nb$_{3}$S$_{6}$, used for single crystal growth afterwards, were synthesized by gas phase method.
Powders of Cr, Nb and S, mixed in a molar ratio of $x_{\rm nominal} : 3 : 6$, were sealed in an evacuated silica tube.
The powder in the silica tube was placed in an electric tube furnace at 1000 $^\circ$C for a week.
The several polycrystalline specimens with $x_{\rm nominal}$ from 1.00 to 1.32 were prepared.
The polycrystals were evaluated by powder X-ray diffraction experiments with Cu $K\alpha$ radiation (Rigaku MultiFlex) and magnetization measurements using a superconducting quantum interference device (SQUID) magnetometer (Quantum Design MPMS-5).

The single crystals were obtained by chemical vapor transport (CVT) technique in a temperature gradient using iodine I$_{2}$ as a transporting agent.
The polycrystalline sample, synthesized with a mixture of Cr, Nb and S in the molar ratio of $x_{\rm nominal} : 3 : 6$, was placed at one end of an evacuated silica tube
and then heated in the electric tube furnace with a temperature gradient for two weeks.
The single crystals were grown at another end of the silica tube located at the lower temperature.
As described in the following, the growth condition affects the amount of Cr intercalation and thus the magnetic behavior.
The molar ratio of Cr, Nb and S in a single crystalline sample was determined by quantitative analysis using electron probe microanalyzer (JEOL JXA-8200S).
The ZAF correction method (Z: atomic number factor, A: absorption factor, F: characteristic fluorescence correction) using standard specimens of each element was carefully applied to the raw data. To examine a locational variation of the molar ratio in the specimen, EPMA data was obtained in ten different positions in each specimen. Then, the averaged values of molar ratio were derived for all the specimens together with the error bar including information of standard deviation between the positions.
The amount of Cr intercalation $x_{\rm EPMA}$ in a crystal was determined by the molar ratio between Cr and Nb.
As for the amount of S, it took almost the ideal molar ratio of Nb and S. Thus, we discuss the defects on the Cr sites in this study.
The magnetic property was evaluated by the magnetization in the presence of small magnetic field orienting in a direction perpendicular to the $c$-axis using SQUID magnetometers (Quantum Design MPMS-5 and MPMS3).

\section{Experimental Results}

\begin{figure}[tb]
\begin{center}
\includegraphics[width=8.5cm]{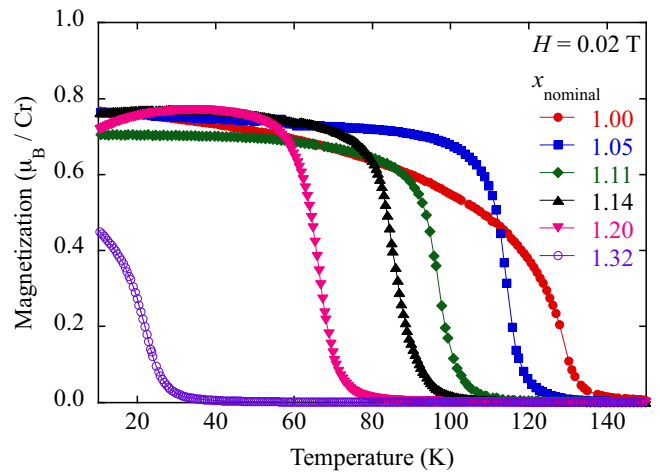}
\end{center}
\caption{Temperature dependence of magnetization obtained by polycrystals of Cr$_{x_{\rm nominal}}$Nb$_{3}$S$_{6}$, synthesized with a mixture of Cr, Nb and S in the molar ratio of $x_{\rm nominal} : 3 : 6$.
}
\label{f-MT-polycrystal}
\end{figure}

The powder X-ray diffraction experiments revealed that the polycrystals with $x_{\rm nominal}$ from 1.00 to 1.32 had no impurity phases since all the diffraction peaks in each sample were indexed to the CrNb$_{3}$S$_{6}$ structure. 
The magnetic property was evaluated by the temperature dependence of magnetization.
As shown in Fig.~\ref{f-MT-polycrystal}, all the specimens exhibit the ferromagnetic response.
The value of $T_{\rm c}$ of the polycrystal with $x_{\rm nominal}$ = 1.00 is 130 K.
$T_{\rm c}$ decreases with increasing $x_{\rm nominal}$, and eventually drops down to 30 K when $x_{\rm nominal}$ becomes 1.32.
However, super high-resolution powder neutron diffraction studies with the $x_{\rm nominal}$ = 1.00 specimen exhibit magnetic satellite peaks as evidence of the CHM order \cite{Kousaka2016}.
Thus, the magnetization measurements with polycrystalline specimens are not useful for identifying the CSL formation in the polycrystals and finding the optimum $x_{\rm nominal}$ for the single crystal growth.

\begin{figure}[tb]
\begin{center}
\includegraphics[width=8.5cm]{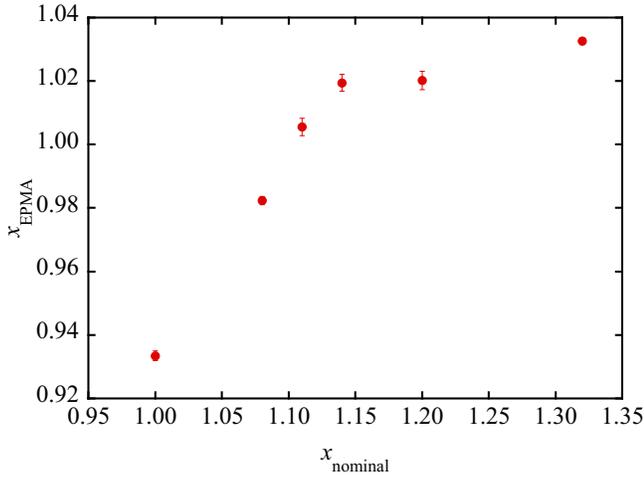}
\end{center}
\caption{
Evaluation of the amount of Cr intercalation $x$ in single crystals of Cr$_{x}$Nb$_{3}$S$_{6}$.
The single crystals were obtained by CVT technique from polycrystalline samples, synthesized with a mixture of Cr, Nb and S in the molar ratio of $x_{\rm nominal} : 3 : 6$.
By using electron probe microanalysis (EPMA), the Cr intercalation of the single crystal $x_{\rm EPMA}$ was directly determined.
}
\label{f-EPMA1}
\end{figure}

As for the single crystal growth, the single crystals with a hexagonal plate shape were obtained in the growth conditions used in the experiments.
The actual amount of Cr intercalation in each specimen was evaluated by EPMA.
It was found that the Cr intercalation $x_{\rm EPMA}$ of the single crystal strongly depends on $x_{\rm nominal}$ of the polycrystals as well as on the growth condition such as the temperature gradient in the furnaces.
Here, to find an appropriate growth condition for obtaining the single crystals of CrNb$_{3}$S$_{6}$,
we show a growth strategy by optimizing $x_{\rm nominal}$ under a fixed experimental condition of the crystal growth.
Figure~\ref{f-EPMA1} shows the amount of Cr intercalation $x_{\rm EPMA}$ of single crystalline Cr$_{x}$Nb$_{3}$S$_{6}$ grown from polycrystals with $x_{\rm nominal}$ under the fixed temperature gradient from 1100 $^\circ$C to 1000 $^\circ$C.
Note that the distance between the highest and lowest temperature locations and the density of I$_{2}$ inside in the silica tube were fixed to be 16 cm long and 5 mg / cm$^3$, respectively.

When the single crystals were grown from polycrystalline samples with $x_{\rm nominal} = 1$, the $x_{\rm EPMA}$ of single crystals was 0.933 (2), indicating the presence of defects in the Cr sites.
$x_{\rm EPMA}$ of the crystals increases in proportion to the $x_{\rm nominal}$ of polycrystals up to $x_{\rm nominal} = 1.14$ and is nearly saturated at 1.019 (3) above it.
Judging from the actual composition determined by EPMA analysis, the single crystal growth should be performed by using polycrsytals with $x_{\rm nominal} \sim 1.1$ so as to reduce the generation of defects on the Cr sites.
Let us note that that the linear correlation between $x_{\rm nominal}$ and $x_{\rm EPMA}$ makes it easy to optimize the crystal growth condition.
On the other hand, in the case of optimizing the temperature setting in the furnace by using the polycrystals with a fixed value of $x_{\rm nominal}$,
it is difficult to find a good combination of many experimental parameters such as temperature values and gradient at high and low temperature locations and the distance between them.
In the present synthesis condition, the obtained crystals exhibit a uniformly distribution of Cr over the specimen because the error bars of the Cr amount in the EPMA analyses were quite low as shown below. Typical values were one hundred times smaller than that reported by Mao {\it et al.} \cite{Mao2022} shown in Table~\ref{tbl-SampleComparison}. In this respect, the values of error bar are likely to reflect the quality of crystals.

\begin{figure}[tb]
\begin{center}
\includegraphics[width=8.5cm]{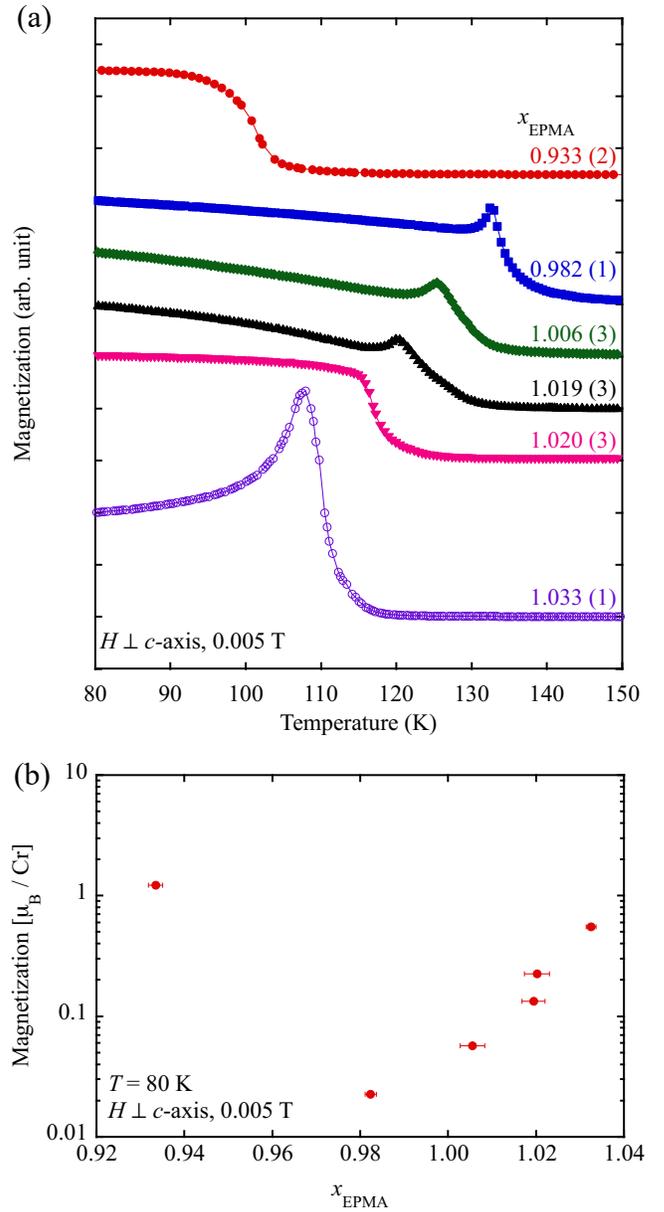}
\end{center}
\caption{(a) Temperature dependence of magnetization obtained by single crystals of Cr$_{x_{\rm EPMA}}$Nb$_{3}$S$_{6}$ in small magnetic field applied perpendicular to the $c$-axis.
Cr intercalation $x_{\rm EPMA}$ in each crystal was determined by electron probe microanalysis (EPMA).
The data, normalized by the magnetization at 80 K in each $M$-$T$ curve, are given in an arbitrary unit.
(b) The magnetization intensity at 80 K obtained by the $M$-$T$ curves in the Cr$_{x_{\rm EPMA}}$Nb$_{3}$S$_{6}$ crystals. The vertical axis is set in a logarithmic scale.
}
\label{f-MT-singlecrystal}
\end{figure}

Figure~\ref{f-MT-singlecrystal}(a)
shows the temperature dependence of magnetization in single crystals with different $x_{\rm EPMA}$.
Here, a small magnetic field is applied in a direction perpendicular to the $c$-axis,
and the amplitude of the magnetization in each specimen was normalized by the magnetization at 80 K.
The single crystals show the CSL or ferromagnetic behavior depending on the values of $x_{\rm EPMA}$.
The presence of chiral helimagnetism could be demonstrated by identifying a sharp peak anomaly of the magnetization as a function of temperature \cite{Kishine2005, Kousaka2009}, as shown in Fig.~\ref{f-CSL-Magnetization}.
$T_{\rm c}$ of the chiral helimagnet crystal is defined at a peak top of the magnetization, while $T_{\rm c}$
of the ferromagnetic order is determined by an extrapolation of the steepest portion of the magnetization toward zero magnetization.
The CSL behavior is observed in the crystals with $x_{\rm EPMA}$ from 0.98 to 1.03.
The single crystals with $x_{\rm EPMA}$ = 0.982 (2) show the CSL response with the highest $T_{\rm c}$ of 133 K.
On the other hand, when the value of $x_{\rm EPMA}$ is a few percent larger or smaller than 1, the ferromagnetic behavior appears with low values of $T_{\rm c}$.
For example, the single crystal with $x_{\rm EPMA}$ = 0.933 (2), obtained from the polycrystal with $x_{\rm nominal} = 1$, exhibits the ferromagnetic behavior with $T_{\rm C} = 102$ K.
This value is much lower than the values of $T_{\rm c}$ for chiral helimagnets.
Note that, as shown in Fig.~\ref{f-MT-singlecrystal}(b),
the amplitude of the magnetization at 80 K in $x_{\rm EPMA}$ = 0.933 (2) is twenty-four times larger than that in $x_{\rm EPMA}$ = 0.982 (2). Such a large difference in the magnetization intensity can be regarded as another criterion to distinguish the ferromagnetic response from the CSL one.
For the absolute distinguishing of the CSL and ferromagnetic phases, further diffractive study using x-ray, electron, and neutrons is required, as described below.

\begin{figure}[tb]
\begin{center}
\includegraphics[width=8.5cm]{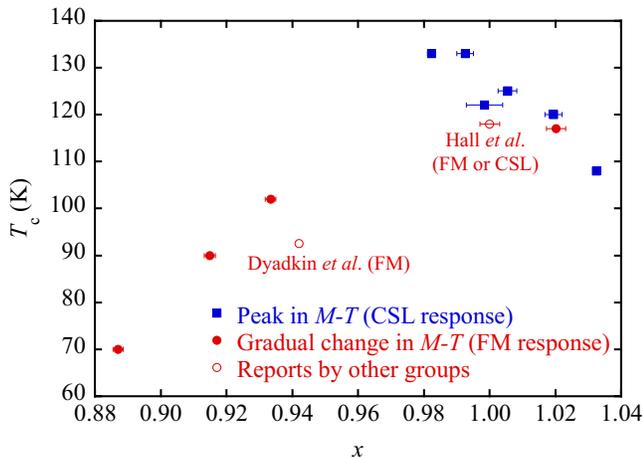}
\end{center}
\caption{
$T_{\rm c}$ versus amount of Cr intercalation $x_{\rm EPMA}$ in single crystals of Cr$_{x_{\rm EPMA}}$Nb$_{3}$S$_{6}$.
Red circles and blue squares represent the transition temperatures,
determined with a presumption of ferromagnetic (FM) and CSL formation respectively,
in the present study.
The open circles correspond to the data by Dyadkin and Hall, in which the ferromagnetic behavior and composition were reported by using single crystal X-ray diffraction and magnetization measurements \cite{Dyadkin2015,Hall2022}.}
\label{f-EPMA2}
\end{figure}

To summarize the experimental results, Fig.~\ref{f-EPMA2} shows a correlation between the Cr intercalation $x_{\rm EPMA}$ and magnetic transition temperature $T_{\rm c}$ in single crystalline Cr$_{x_{\rm EPMA}}$Nb$_{3}$S$_{6}$.
Note that most of the present data, denoted by circles and squares, were taken from the crystals grown in the same experimental condition with the fixed temperature gradient.
The data at $x =$ 0.887 (2), 0.915 (2), 0.993 (2) and 0.995 (6) were obtained in the crystals in the different experimental condition.
The CSL behavior is observed in a narrow region of $x_{\rm EPMA}$ from 0.98 to 1.03. $T_{\rm c}$ takes the highest value at 0.98 and 0.99.
When $x_{\rm EPMA}$ is larger or smaller than $0.98 \sim 0.99$, $T_{\rm c}$ is likely to decrease monotonously.

\section{Discussion}
In growing single crystals of CrNb$_{3}$S$_{6}$, the nominal amount of Cr $x_{\rm nominal}$ is one of the most important and controllable parameters.
As demonstrated in the present study, the Cr intercalation $x$ of single crystals is smaller than $x_{\rm nominal}$ in polycrystals used for the crystal growth.
Resultantly, Cr-defected crystals with $x < 1$ are synthesized without an optimization of the growth condition.
In the present study, we found that the Cr intercalation $x$ of single crystals is proportional to $x_{\rm nominal}$ up to some amount of $x_{\rm nominal}$ (up to $x \sim 1.02$ in our experimental condition).
Thus, the optimization of $x_{\rm nominal}$ enables us to obtain CrNb$_{3}$S$_{6}$ single crystals without defects on the Cr sites.

Note that the magnetic properties are influenced by the Cr intercalation $x$ of single crystals as well.
As shown in Fig.~\ref{f-EPMA2}, the highest $T_{\rm c}$ of 133 K for chiral helimagnetism is obtained for $x = 0.98$ and 0.99.
A deviation of the $x$ values from the optimum one gradually decreases the value of $T_{\rm c}$.
When $x$ is three or four percent smaller than the optimized $x$, the single crystals do not show chiral helimagnetism but exhibit the ferromagnetism.
When increasing the $x$ value, $T_{\rm c}$ for chiral helimagnetism decreases and the ferromagnetism appears in some crystals.
This behavior reminds us of a dome-shaped distribution of the $T_{\rm c}$ of high-$T_{\rm c}$ oxide superconductors against a carrier doping.
The value of $T_{\rm c}$ is proportional to the strength of $J_{\perp}$, which stabilizes the coherency of the chiral magnetic order in CrNb$_{3}$S$_{6}$ \cite{Shinozaki2016}.
In this sense, the amplitude of $J_{\perp}$ and  the CSL robustness can be controlled by the Cr intercalation $x$.

The CSL behavior was observed in the crystals with an excessive intercalation of Cr by 3 percent.
The magnetic properties in an over-intercalated regime would be worth investigation in terms of the robustness of the CSL formation.
In the present experimental condition, the Cr intercalation $x$ was nearly saturated at around 1.02, as shown in Fig.~\ref{f-EPMA1}.
However, the polycrystalline powders in Fig.~\ref{f-MT-polycrystal} show a gradual decrease of $T_{\rm c}$ for the $x_{\rm nominal}$ values from 1.00 to 1.32.
Thus, there is a possibility of obtaining single crystals with such over intercalated Cr ions.
We stress that small angle neutron scattering experiments will directly probe the ferromagnetic or chiral helimagnetic magnetic structure of such single crystals with a variety of the intercalation amount.

Now we know that the single crystals of CrNb$_{3}$S$_{6}$ with smaller values of $T_{\rm c}$ have a deficient or excessive intercalation of the Cr ions. Namely, we can estimate the amount of Cr intercalation $x$ simply by measuring the magnetization.
In this respect, the magnetization gives us an important information to evaluate the quality of the obtained single crystals.
To find the growth condition for obtaining high-quality single crystals, the nominal amount of Cr should be optimally parameterized with the fixed experimental condition regarding the furnace setting.
Similar idea can be applicable for growing single crystals of other kinds of $TM_{3}$S$_{6}$ compounds.

As listed in Table~\ref{tbl-SampleComparison}, some studies have made a discussion on the deficiency of ions in CrNb$_{3}$S$_{6}$.
Mao {\it et al.} pointed out the influence of defects of sulfur (S) elements against a drastic reduction of $T_{\rm c}$ \cite{Mao2022}.
However, as shown in Table~\ref{tbl-SampleComparison}, in their analysis using one particular single crystal that showed the ferromagnetic response at 56 K,
they obtained x = 0.99 (12) within one sigma error, (thus $x$ ranges from 0.87 to 1.11). Based on the results in Fig.~\ref{f-EPMA2}, $T_{\rm c}$ can be less than 60 K for the crystal with $x$ = 0.87.
Thus, such a low $T_{\rm c}$ can be ascribed to the defects of the Cr ions even within the experimental error bar.
To avoid misleading conclusions, systematic examination is required using several crystals with different amount of the defects of the target elements.

Han {\it et al.} reported that their single crystal of CrNb$_{3}$S$_{6}$ had the defects of Cr by 12.5 percent \cite{Han2017}.
The temperature dependence of magnetization showed the ferromagnetic response at 125 K, which is inconsistent with the present results in Fig.~\ref{f-EPMA2}.
Since there was no description on the experimental error of the defect evaluation, the data might be taken at only one location in the crystal.
Interestingly, their crystal exhibited an indication of the CSL formation with much lower $H_{\rm c}$ in the magnetic field dependence of the magnetization.

The correlation between the Cr intercalation $x$ and $T_{\rm c}$, shown in Fig.~\ref{f-EPMA2}, is in good agreement with the results obtained by single crystal X-ray diffraction \cite{Dyadkin2015,Hall2022}.
Dyadkin {\it et al.} reported that a ferromagnetic single crystal of Cr$_{x}$Nb$_{3}$S$_{6}$ with $T_{\rm c}$ = 92 K
had defects in the Cr sites.
For this crystal, the crystal structure analysis gave 0.942 for the $x$ value \cite{Dyadkin2015}.
As shown in Fig.~\ref{f-EPMA2}, our analysis reveals that the crystals with $T_{\rm c}$ = 92 K have about 0.92 for $x$, which roughly reproduces the results by Dyadkin {\it et al.}

Hall {\it et al.} reported $x$ = 1.00 (3) for the Cr$_{x}$Nb$_{3}$S$_{6}$ single crystal with $T_{\rm c}$ of approximately 118 K\cite{Hall2022}.
The crystal structure analysis told that 95.2 (2) percent of Cr occupied Wyckoff $2c$ position, while the remaining Cr occupied Wyckoff $2b$ position.
Although the total amount of the Cr intercalation took an ideal value, the temperature dependence of magnetization showed the ferromagnetism rather than the CSL formation.
Interestingly, the field dependence of magnetization was not simply regarded to be of ferromagnetic origin, although $H_{\rm c}$ was 8 times smaller than the typical value shown in Fig.~\ref{f-CSL-Magnetization}(a).

Similar examples of the single crystals can be found in Fig.~\ref{f-EPMA2}.
The crystals with $x$ = 1.019 (3) and 1.033 (1) exhibits the CSL behavior, while the one with $x$ = 1.020 (3) shows the ferromagnetic response.
The latter is quite similar to the crystal examined by Hall {\it et al.}
Even though the crystals mentioned above have almost the same amount of the Cr intercalation $x$,
the ferromagnetic behavior appears in a particular piece of the single crystal, which may be caused by a delicate balance of the locations of the Cr ions in the single crystals.
Namely, the deficiency of the Cr ions on $2c$ site leads to the additional intercalation on $2b$ site.
Such disorders on the Cr sites may decrease the amplitude of DM interaction significantly, which could be monitored by the value $H_{\rm c}$ since it is proportional to $D^{2}/J_{\parallel}$.
In this connection, it would be interesting to investigate the relationship between the intercalation ratio between $2b$ and $2c$ sites and $H_{\rm c}$ in addition to the evaluation of $T_{\rm c}$.

In order to discuss the dome-shaped behavior of $T_{\rm c}$ in detail, additional experiments will be necessary.
A similar behavior is exhibited by another transition metal sulfides such as ferromagnetic Fe-intercalated tantalum sulfide, and is discussed from the viewpoint of oscillatory characteristics of Ruderman-Kittel-Kasuya-Yosida (RKKY) interaction \cite{Xie2022}.
The Cr-Cr distance determined by single crystal X-ray diffraction measurements and the carrier concentration by Hall measurements will be helpful for understanding the mechanism of the dome-shaped behavior.
In this paper, the ferromagnetic or chiral helimagnetic order is estimated by the existence of the sharp anomaly around $T_{\rm c}$.
Such magnetic structures can be distinguished by the existence of magnetic satellite peaks in neutron diffraction experiments. Regarding $T_{\rm c}$, a precise value can be determined by Arrott plot analysis of the field dependence of magnetization \cite{Ghimire2013} as well as by neutron diffraction and heat capacity measurements.
In the present study, most crystals grown with $x > 1$ reveals the CSL behavior, while a particular crystal with $x$ = 1.020 (3) shows the ferromagnetic response.
This exception may be caused by disorders on the Cr sites, the presence of which can be probed by single crystal X-ray diffraction measurements, as observed in a ferromagnetic CrNb$_{3}$Se$_{6}$ \cite{Gubkin2016}.

\section{Conclusion and Future Perspectives}

We show how to find the appropriate growth condition of a chiral helimagnet Cr$_{x}$Nb$_{3}$S$_{6}$ by optimizing the nominal amount of Cr in the crystal growth.
As shown in Fig.~\ref{f-EPMA1}, the actual Cr intercalation $x$ in single crystals is smaller than the nominal amount of Cr $x_{\rm nominal}$ of polycrystals used in the crystal growth.
$x$ dramatically affects the magnetic properties and has a correlation with $T_{\rm c}$.
Therefore, the amount of Cr intercalation is able to be estimated only by the magnetization measurements.
As shown in Fig.~\ref{f-EPMA2}, the single crystals with $x$ being 0.98 and 0.99 shows the CSL behavior with the highest $T_{\rm c}$ of 133 K.
Thus, the intercalation-controlled single crystals were successfully obtained by using the present approach.
Only several percent of defects of the Cr ions prevent the CSL formation and alternatively stabilize the ferromagnetic order probably because of the reduction of the strength of the DM interaction.
In this sense, it is very important to check the amount of Cr intercalation $x$ in the crystals.

$TM_{3}$S$_{6}$ compounds will be attracting more attention because of the studies on the CSL formation recently made in CrTa$_{3}$S$_{6}$ \cite{Kousaka2016, Zhang2021, Obeysekera2021}.
The strategy for the crystal growth demonstrated for CrNb$_{3}$S$_{6}$ in the present study can be applied to synthesizing other $TM_{3}$S$_{6}$ compounds. 
The synthesis strategy can be categorized to two cases.
The first case is that the compound has a correlation between the amount of $T$ intercalation and $T_{\rm c}$, which is the same situation as that for CrNb$_{3}$S$_{6}$.
Even when the intercalation amount is smaller or larger than the ideal value, the crystal keeps the $TM_{3}$S$_{6}$ phase.
However, a deficient or excessive intercalation affects the physical properties.
For example, such a tendency can be found in the $T_{\rm c}$ distribution as a function of the intercalation amount for CrNb$_{3}$S$_{6}$, as shown in Fig.~\ref{f-EPMA2}. Fortunately, in this case, the growth condition can be optimized by choosing the nominal composition so as to give the highest $T_{\rm c}$.

The second case is that the mixture of $TM_{3}$S$_{6}$ and $TM_{4}$S$_{8}$ is synthesized in the crystal.
When the intercalation amount is smaller than the ideal, the deficient amount of $T$ does not decrease $T_{c}$, but stabilizes some amount of $TM_{4}$S$_{8}$ phase.
For example, MnNb$_{3}$S$_{6}$ and MnNb$_{4}$S$_{8}$ have different $T_{\rm c}$ at 40 K and 100 K \cite{Parkin1980a, Kousaka2009,Onuki1986}, respectively.
The magnetization of the MnNb$_{3}$S$_{6}$ crystal frequently shows two-step transitions at both temperatures \cite{Hall2022}.
In such cases, the volume fraction between MnNb$_{3}$S$_{6}$ and MnNb$_{4}$S$_{8}$ can be determined by the temperature dependence of magnetization.
Then, the actual amount of Mn ions
can be estimated by the volume fraction as well.
The remaining task is the optimization of the growth condition so as to eliminate the $TM_{4}$S$_{8}$ phase in the crystal.

The next challenge in synthesizing $TM_{3}$S$_{6}$ compounds will be a development of the method of enantiopure crystal growth. 
Non-reciprocal magneto-chiral response and CISS effect are recently observed in CrNb$_{3}$S$_{6}$ \cite{Aoki2019,Inui2020}.
These findings shed light on an importance of synthesizing mono-chiral crystals because these phenomena are coupled with crystallographic chirality of the compounds. 
However, it is still difficult to stabilize only left- or right-handed chiral crystal structure in synthesizing inorganic chiral materials
because it is inevitable to prevent the formation of racemic-twinned crystals, having the left- and right-handed crystalline domains within a specimen.
Indeed, the CrNb$_{3}$S$_{6}$ crystals obtained by CVT method form the racemic-twinned grains \cite{Inui2020}.
In some inorganic compounds such as NaClO$_{3}$, $T$Si, CsCuCl$_{3}$ and YbNi$_{3}$Al$_{9}$, a recent progress of growth techniques makes it possible to obtain the enantiopure crystals with a desired handedness \cite{Kondepudi1990, Dyadkin2011, Kousaka2014, Kousaka2017, Nakamura2020, Kousaka2022}.
However, such methods are not applicable in the synthesis of $TM_{3}$S$_{6}$ and thus a new method will be required to obtain enantiopure crystals.
These challenges will be a key factor to observe new physical properties coupled with the handedness of the crystals.

\begin{acknowledgments}
This work was supported by JSPS KAKENHI Grant Numbers 15H05885, 19KK0070, JP19H05822, JP19H05826 and 20H02642.
\end{acknowledgments}

\section*{Data Availability Statement}
The data that support the findings of this study are available from the corresponding author upon reasonable request.

%\bibliography{apssamp}% Produces the bibliography via BibTeX.

\end{document}